\documentclass[conference]{IEEEtran}
\IEEEoverridecommandlockouts
\usepackage{graphicx, tikz, cite}
\usepackage{amsmath}
\usepackage{amsthm}
\usepackage{amssymb}
\usepackage{epstopdf}
\usepackage{subfigure}
\usepackage{color}
\usepackage{graphicx, psfrag, minibox, mathrsfs, bigints, stfloats, color, upgreek, gensymb, textcomp, euscript, calligra, xcolor, pict2e}
\date{}
\usepackage{algorithmicx}
\usepackage{algpseudocode}
\usepackage[ruled]{algorithm2e}

\usepackage{setspace}
\usepackage{float}
\usepackage[font=scriptsize]{caption}
\usepackage{enumerate}
\usepackage{eso-pic}
\definecolor{red}{rgb}{0.8,0.1,0.1}
\definecolor{blue}{rgb}{0,0,1}

\begin{document}
\title{Power Adaptation for Distributed Detection in Energy Harvesting WSNs with Finite-Capacity Battery}
\author{\IEEEauthorblockN{Ghazaleh Ardeshiri,
		Hassan Yazdani,
		Azadeh Vosoughi~\IEEEmembership{Senior Member,~IEEE}}
	\IEEEauthorblockA{University of Central Florida\\  Email:gh.ardeshiri@knights.ucf.edu, h.yazdani@knights.ucf.edu, azadeh@ucf.edu} }
\maketitle
\begin{abstract}
We consider a wireless sensor network, consisting of $N$ 
heterogeneous sensors and a fusion center (FC), that is tasked with solving a binary distributed detection problem. Each sensor is capable of harvesting randomly arrived energy and storing it in a finite capacity battery. Sensors are informed of their fading channel states, via a bandwidth-limited feedback channel from the FC. Each sensor has the knowledge of its current battery state and its channel state (quantized channel gain). Our goal is to study how sensors should choose their transmit powers such that $J$-divergence of the received signal densities under two hypotheses at the FC is maximized, subject to certain (battery and power) constraints. We derive the optimal power map, which depends on the energy arrival rate, the battery capacity, and the battery states probabilities at the steady state. Using the optimal power map, each sensor optimally adapts its transmit power, based on its battery state and its channel state. Our simulation results demonstrate the performance of our proposed power adaptation scheme for different system parameters. 
\end{abstract}
\IEEEpeerreviewmaketitle

\section{Introduction}
A wireless sensor network (WSN), consisting of a network of sensors with embedded capabilities of sensing, computation, and communication, is typically used to sense and collect data for a wide range of applications \cite{Sudevalayam}. Traditionally, a WSN is composed of sensor nodes powered by non-rechargeable batteries with limited energy storage capacities. As a result, a WSN can only function for a limited time \cite{Mao}.
Recently, energy harvesting (the technology of harnessing energy from renewable resources in ambient environment such as solar, wind, and geothermal energy)
has attracted much attention \cite{Ardeshiri}. 
Utilizing harvesting technology in WSNs can pave the way to building a self-sustainable system with a lifetime that is not limited by the lifetime of the conventional batteries \cite{Tarighati}. 
Unlike traditional battery-powered systems, where transmission is often subject to a constant power constraint, the energy available to an energy harvesting system is modeled as a random process \cite{Zhao}. For transmitters that are powered by energy harvesters, unlike conventional communication devices that are subject only to a power constraint or a sum energy constraint, they, in addition, subject to other energy harvesting constraints \cite{Ho,J1}. 
\par In this paper, we consider the distributed detection of a known signal using a WSN with $N$ energy harvesting sensors and a fusion center (FC). Each sensor makes a noisy observation and has a battery with a finite capacity.
Each sensor makes a local decision based on its own observation. Sensors send their local decisions to the FC over orthogonal channels, that are subject to fading and additive white Gaussian noise (AWGN). Assuming the knowledge of channel gains at the FC, the FC feeds back the quantized channel gains to the sensors via a bandwidth-limited feedback channel. Given its battery state and the quantized channel gain, each sensor adjusts its transmit power accordingly.  Our goal is to study how each sensor should optimally adapt its transmit power, such that the detection performance metric at the FC is optimized, subject to certain battery related constraints. 
%
%
\begin{figure}[!t]
	\centering
	\includegraphics[scale=.335]{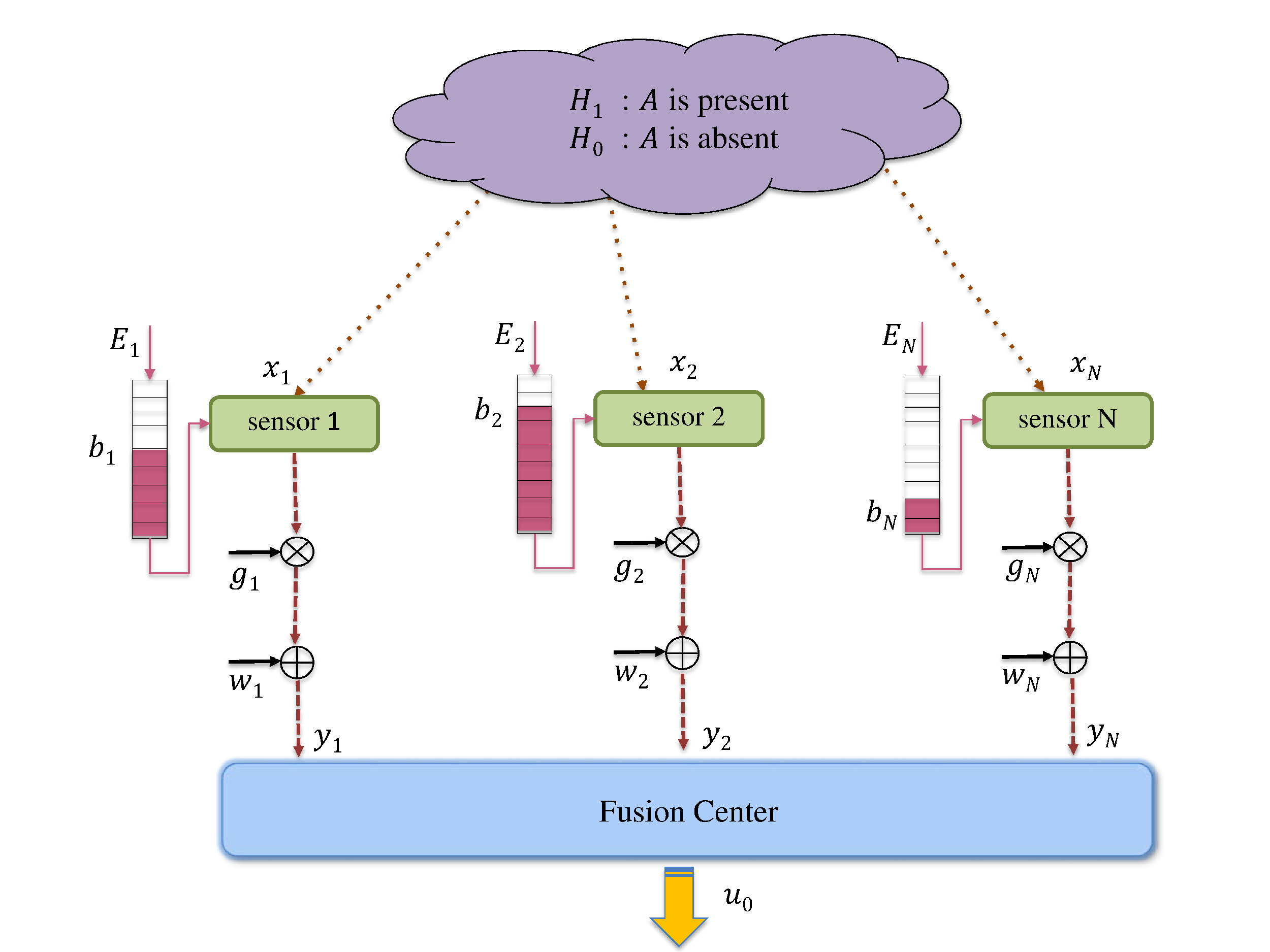}
	\caption{Our System model during one observation period.}
	\label{system model}
	\vspace{-2mm}
\end{figure}
We choose $J$-divergence as our detection performance metric, motivated by the fact that it is a widely used metric for detection systems, since it provides a lower bound on the detection error probability. Furthermore, it allows us to derive analytically tractable results in our study. In particular, we find closed-form solutions for the optimal transmit powers at the sensors that specify how each sensor should choose its transmit power, according to its battery state and its channel state information.
\\The paper organization follows: Section \ref{sym} describes our system model, including  transmission model, energy harvesting system, and the battery state model. Section \ref{opt} derives an approximate expression for the $J$-divergence. Section \ref{XYZ} formalizes our optimal power adaptation problem and provides its solution.
 Section \ref{sim} illustrates our numerical results and Section \ref{conclu} includes our concluding remarks. 
\section{System Model}\label{sym}
In this section we describe the distributed detection problem and we address our system setup, including transmission model and the battery state model (based on the consumed energy for transmitting local decisions and the randomly harvested energy).
\subsection{Distributed Detection Problem}
We address a binary distributed detection problem in a WSN, consisting of $N$ sensors and a FC. Sensors are deployed to distinguish between two hypotheses $\mathcal{H} =\left\{\mathcal{H}_0, \mathcal{H}_1\right\}$, with probabilities $\Pi_0\!=\!P(\mathcal{H}_0)$ and $\Pi_1\!=\!P(\mathcal{H}_1)\!=\!1\!-\!\Pi_0$, where $\mathcal{H}_0$ and $\mathcal{H}_1$ represent absence or presence of known scalar signal ${\cal A}$, respectively (see Fig.~\ref{system model}). Let $x_n$ denote the local observation at sensor $n$ during an observation period. We assume the following signal model
\begin{equation}\label{xk}
\mathcal{H}_{1}: x_{n} ={\cal A}+v_{n}, ~~~~~\mathcal{H}_{0}: x_{n}=v_{n},
\end{equation}
where $v_{n}$ is the additive observation noise. We assume $v_{n} \! \sim \! {\cal N}(0,\sigma_{v_{n}}^2)$ and
all observation noises are independent over time and among $N$ sensors. 
Sensor $n$ makes a local binary decision $u_{n}$, independent of other sensors, according to a certain local decision rule based on $x_{n}$ only.  
Let $\Gamma_n(.)$ denote the local decision rule for sensor $n$. The local decision, $u_n$ is
\begin{equation}\label{u_n}
u_{n} = \Gamma_n(x_n)=
\begin{cases}
1, &~~~ \text{decide}~{\cal H}_1\\
0, &~~~ \text{decide}~{\cal H}_0
\end{cases}
\end{equation}
Let $P_{f_n}$ and $P_{d_n}$ denote, respectively, the false alarm and detection probabilities at sensor $n$, i.e., $P_{f_n} \!= \! \Pr(u_n\! =\! 1 |\mathcal{H}_{0})$ and $P_{d_n} \!= \!\Pr(u_n\!= \!1 |\mathcal{H}_{1})$.
Sensors send their local decisions to the FC over orthogonal channels that are subject to fading and AWGN.  The received signal at the FC from sensor $n$ is
\begin{equation}\label{y_n}
   y_{n}=\sqrt{g_n}a_n u_{n}+w_{n} ~~~~\text{for}~n=1,\dots,N
\end{equation}
 where $g_n$ is the exponential fading channel gain corresponding to sensor $n$ with parameter $\gamma_{g_{n}}$ and $w_n \sim {\cal N} (0, \sigma_{w_n}^2).$ Also, $a_n$ is the amplitude of the signal transmitted by sensor $n$, and hence $P_n=a_n^2$ is the transmitted power of sensor $n$ corresponding to its local decision $u_n\!=\!1$. 
 Given the knowledge of channel gains $g_n$'s, 
 the FC quantizes $g_n$'s and sends the quantized gains to the sensors through a limited feedback channel. Hence, sensors can optimally adjust their transmit powers $P_n$'s according to their channel state information. In particular, suppose the FC partitions the set of positive real numbers into $L$ disjoint intervals for sensor $n$, denoted as $\mathcal{I}_{n,1}, \dots, \mathcal{I}_{n,L}$. For sensor $n$ these quantization intervals are determined by the quantization thresholds $\{ \mu_{n,l}\}_{ l=1}^L$, where $0\! = \mu_{n,0}\! < \mu_{n,1}\! < \dots \!< \mu_{n,L+1}\! = \infty$. 
 In other words $\mathcal{I}_{n,l}=[\mu_{n,l}, \mu_{n,l+1})$ for $l=0,\ldots,L$. The channel gain quantization rule at the FC for sensor $n$ follows: if $g_n \in \mathcal{I}_{n,l}$ then $g_n$ is quantized to $\mu_{n,l}$. We assume that the channel coherence time is larger than an observation period and hence the channel gains are unchanged during this time. We define $\pi_{n,l}=\Pr(\mu_{n,l}\leqslant g_n<\mu_{n,l+1})$, which can be found based on the distribution of fading model in terms of the thresholds $\mu_{n,l}$ and $\mu_{n,l+1}$.
 Let $\boldsymbol{y}=[y_1,y_2, \ldots ,y_N]$ denote the vector that includes the received signals at the FC from all sensors. The FC applies its fusion rule $\Gamma_0(.)$ to $\boldsymbol{y}$ and obtains a global decision $u_0=\Gamma_0(\boldsymbol{y})$ where $u_0 \in \{0,1\}$.
The conditional probability density functions (pdfs) of $\boldsymbol{y}$ given the two hypotheses are
\begin{align}\label{y/H}
f(\boldsymbol{y}|\mathcal{H}_{i})&= \prod_{n=1}^N f(y_n|u_n, \mathcal{H}_i)\Pr(u_n|\mathcal{H}_i) \nonumber\\
&=\prod_{n=1}^N \underbrace{f(y_n|u_n)\Pr(u_n|\mathcal{H}_i)}_{=f(y_n|\mathcal{H}_i)} ~~~~~~\text{for}~i=0,1
\end{align}
 Note given $u_n$, $y_n$ and $\mathcal{H}_{i}$ are independent and hence $f\left(y_{n}|u_n,\mathcal{H}_{i}\right)\!=\!f\left(y_{n}|u_n\right)$ for $i\!=\!0,1$. Also, given $u_n$ and $g_n$, $y_n$ is Gaussian. In particular,  $y_{n}|_{u_{n}=0}\sim {\cal N}\left(0,\sigma_{w_n}^2\right)$ and $y_{n}|_{u_{n}=1}\sim {\cal N}\left(\sqrt{g_n}a_n,\sigma_{w_n}^2\right)$.
\subsection{Battery State Model}\label{battery_st}
%
\begin{figure}[!t]
\centering
\scalebox{0.7}{
\begin{tikzpicture}
\small
\draw [thick, fill=pink!100] (-2,0) rectangle (0.8,0.8);
\node at (-0.5,0.4) {Slot $t$};
\draw [thick, fill=pink!100] (0.8,0) rectangle (3.6,0.8);
\node at (2.2,0.4) {Slot $t+1$};
\draw [-] (-2,0) -- (-2,-0.25);
\draw [-] (0.8,0) -- (0.8,-0.25);
\draw [-] (3.6,0) -- (3.6,-0.25);
\draw [<->] (-2,-0.15) -- (0.8,-0.15);
\draw [<->] (0.8,-0.15) -- (3.6,-0.15);
\draw [->] (-2,1.3) -- (-2,0.8);
\draw [->] (0.8,1.3) -- (0.8,0.8);
\draw [->] (3.6,1.3) -- (3.6,0.8);
\node at (-0.5,-0.4) {$T_s$};
\node at (2.2,-0.4) {$T_s$};
\node at (-2,1.6) {$E^{t}_n$};
\node at (0.8,1.6) {$E^{t+1}_n$};
\node at (3.6,1.6) {$E^{t+2}_n$};
\end{tikzpicture}
}
\caption{Frame structure.} 
\label{Frame}
\vspace{-5mm}
\end{figure}
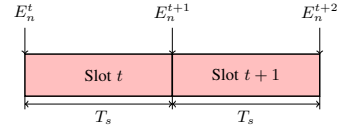
We assume each sensor is able to harvest randomly arrived energy from the environment and stores it in a battery. This battery has the capacity of storing at most $K$ units (cells) of energy, where each unit is equivalent to $e_u$ Jules. 
 Therefore, the battery capacity measured in Jules is equal to $K e_u$. When $k$ units of the battery is fully charged we say that the battery is at state $k$.
 Let $b_n^t = k $, $k = 0, 1,\dots, K$  denote the energy state information (ESI) of sensor $n$ at the beginning of slot $t$. Note that $b_n^t = 0$ represents the empty battery level, and $b_n^t = K$ represents the full battery level. 
 Suppose sensors use a frame with duration $T_s$ (see Fig. \ref{Frame})
and $E_n^t$ denotes the arrival energy during slot $t$ at sensor $n$.
The energy arrives randomly in each slot with a fixed energy arrival rate $\gamma_e$. The energy arrival process $E_n^t$ is typically modeled as a sequence of independent and identically distributed (i.i.d) random variables with an exponential distribution \cite{Zhang,Altinel}.
Hence, the cumulative distribution function (CDF) of $E_n^t$ is $F_{E_n} (x) \! = \! (1-e^{\frac{-x}{\gamma_e}}) u(x)$.
The battery harvests energy and stores it in $\beta_n^t = \lceil E_n^t/e_u\rceil$ units. We assume that the energy harvested at slot $t$ is immediately available in slot $t + 1$ for data transmission.
\par Suppose $\alpha_{n,l,k}^t$ denotes the number of energy units consumed at slot $t$ when $g_n \in \mathcal{I}_{n,l}$ and the battery is at state $k$ ($b_n^t \! = \! k$) for sensor $n$. Also, we assume that the energy consumed for sensing is negligible. The battery state in the next time slot (slot $t+1$) depends on the present system state (slot $t$) and the action taken in slot $t$ (whether the sensor local decision $u_n=0$ or $u_n=1$). If sensor $n$ decides ${\cal H}_0$, then the battery state in the next time slot is
\begin{equation}
    b_n^{t+1} = \min\big\{[b_n^t + \beta_n^t]^+,K\big\}
\end{equation}
If sensor $n$ decides ${\cal H}_1$, then the battery state in the next time slot is
\begin{equation}
    b_n^{t+1} = \min\big\{[b_n^t + \beta_n^t-\alpha_{n,l,k}^t]^+,K \big \}
\end{equation}
where $[x]^+=\max \{0,x\}$. We define $\psi^t_{n,k} = \Pr(b_n^t=k )$ as the probability that the battery state of sensor $n$ at slot $t$ is equal to $k$. Note that $\psi_{n,k}^t$ depends on the battery state at slot $t-1$, the harvested energy at slot $t-1$, and the transmit power at slot $t-1$ if sensor $n$ decides $\mathcal{H}_1$. Therefore, we can write $\psi^{t}_{n,k}$ as \eqref{prob_battery}.
\begin{figure*}
\begin{subequations}\label{prob_battery}
\begin{align}
    &\psi_{n,K}^{t}\!=\! \Pr(b_n^{t}\! =\! K)\!=\! \sum^L_{l=1}\sum_{k=0}^K \pi_{n,l}\; \psi^{t-1}_{n,k}\Big[\Pi_0\Pr\big( k +\beta^{t-1}_n \geq K\big)+\Pi_1\Pr\big(k +\beta^{t-1}_n -\alpha^{t-1}_{n,l,k} \geq K \big)\Big]\\
    &\psi_{n,0}^{t}\!=\! \Pr(b_n^{t}\! = \!0)\!=\! \sum^L_{l=1}\sum_{k=0}^K \pi_{n,l}\; \psi^{t-1}_{n,k}\Big[\Pi_0\Pr\big(k +\beta^{t-1}_n \leq 0\big)+\Pi_1\Pr\big(k +\beta^{t-1}_n -\alpha^{t-1}_{n,l,k} \leq 0 \big)\Big]\\
   &\psi_{n,j}^{t} \!= \!\Pr(b_n^{t}\! =\! j )\!=\! \sum^L_{l=1}\sum_{k=0}^K \pi_{n,l}\; \psi^{t-1}_{n,k}\Big[\Pi_0\Pr\big(k +\beta^{t-1}_n\!=\! j\big)+\Pi_1\Pr\big(k +\beta^{t-1}_n -\alpha^{t-1}_{n,l,k} \!=\! j\big)\Big],~~\text{for}~ 1\!\leq\! j \!\leq \!K\!-\!1
\end{align}
\end{subequations}
\hrulefill
\vspace{-3mm}
\end{figure*}
\section{Characterization of $J$-Divergence}\label{opt}
Given our system model, our goal is to optimize the transmit powers $P_n$'s  for all sensors, such that the detection performance at the FC is optimized. Natural choices for the detection performance metric are detection and false alarm probabilities (or error probability) corresponding to the global decision $u_0$ at the FC. However, finding closed-form expressions of these probabilities, even for the centralized detection, is very difficult.  We choose one of the distance related bounds of the Ali-Silvey class of distance measures, specifically, the $J$-divergence, as our detection performance metric \cite{vin}. Our choice is motivated by the facts that (i) it is a widely used metric for evaluating detection system performance \cite{Guo}, \cite{Sami}, since it provides a lower bound on the detection error probability, (ii) it is closely related to other types of detection performance metric, including  the asymptotic relative efficiency (ARE). Given that sensor $n$ knows its quantized channel gain and the state of its battery, we study how each sensor should optimally adapt its transmit power, such that the $J$-divergence at the FC is maximized, under certain constraints related to the network power and individual batteries (will be discussed in details in Section \ref{XYZ}). Our proposed transmit power adaptation can be implemented in a distributed fashion, i.e., each sensor adapts its transmit power according to its locally available information about its fading channel gain and its battery state. 
The $J$-divergence between two probability densities, denoted as $\rho_1$ and $\rho_0$, is defined as
\begin{equation}\label{first_j}
    J(\rho_1,\rho_0) = D(\rho_1||\rho_0)+D(\rho_0||\rho_1)
    \vspace{-1mm}
\end{equation}
where $D(\rho_1||\rho_0)$ is the non-symmetric Kullback-Leibler (KL) distance between $\rho_1$ and $\rho_0$. The KL distances  $D(\rho_1||\rho_0)$ and $D(\rho_0||\rho_1)$ are defined as
\begin{equation}\label{kl}
    D(\rho_i||\rho_j) = \int \log \left(\frac{\rho_i}{\rho_j}\right)\rho_i.
\end{equation}
Using \eqref{kl} we can write  the $J$-divergence 
as
\begin{align}\label{j_j}
J\big (f(\boldsymbol{y} & |{\cal H}_1),f(\boldsymbol{y}| {\cal H}_0)\big)= \\ 
&\sum_{n=1}^N \underbrace{ \!\int_{y_n} \!\!\! \Big [f(y_n| {\cal H}_1)\! - \! f(y_n| {\cal H}_0)\Big ] {\rm log}{ \frac{f(y_n| {\cal H}_1)} {f(y_n| {\cal H}_0)}} \;dy_n }_{= \; J_n(f(y_n|{\cal H}_1),f(y_n|{\cal H}_0))}\!. \nonumber
\end{align}
As pointed out in \cite{vin}, the conditional pdfs $f(y_n|{\cal H}_i)$ are Gaussian mixtures. Unfortunately, the $J$-divergence between two Gaussian mixture densities does not have a general
closed-form expression.
Similar to \cite{vin}, we approximate the $J$-divergence between two Gaussian mixture
densities by the $J$-divergence between two Gaussian densities $f^G(y_n|{\cal H}_i) \sim {\cal N}(m_{n,{\cal H}_i},\Sigma_{n,{\cal H}_i})$, where the parameters $m_{n,{\cal H}_i}$ and $\Sigma_{n,{{\cal H}_i}}$ of the approximate distributions are obtained from matching the first and second order moments of the actual and the approximate distributions. For our problem setup, one can verify that the parameters $m_{n,{\cal H}_i}$ and $\Sigma_{n,{{\cal H}_i}}$ become
\begin{align*}
    m_{n,{\cal H}_0}=\sqrt{P_n g_n}P_{f_n}\nonumber,~~~ \Sigma_{n,{{\cal H}_0}}\!=\!P_n g_n P_{f_n}(1\!-\!P_{f_n})\!+\!\sigma_{w_n}^2\\ 
m_{n,{\cal H}_1}\!=\!\sqrt{P_n g_n}P_{d_n},~~~ \Sigma_{n,{{\cal H}_1}}\!=\!P_n g_n P_{d_n}(1\!-\!P_{d_n})\!+\!\sigma_{w_n}^2
\end{align*}
The $J$-divergence between two Gaussian densities, represented as $J_n\big(f^G(y_n|{\mathcal H}_1),f^G(y_n|{\cal H}_0)\big)$, in terms of their means and variances is \cite{vin}
\begin{align}\label{g_J}
    & J_n\big(f^G(y_n|{\cal H}_1),f^G(y_n| {\cal H}_0)\big)= \\
     &~~ \frac{\Sigma_{n,{{\cal H}_1}}\!+\!( m_{n,{\cal H}_1}\!-\! m_{n,{\cal H}_0})^2}{\Sigma_{n,{{\cal H}_0}}}
     +\frac{\Sigma_{n,{{\cal H}_0}}\!+\!( m_{n,{\cal H}_0}\!-\! m_{n,{\cal H}_1})^2}{\Sigma_{n,{{\cal H}_1}}}. \nonumber
\end{align}
Substituting $m_{n,{\cal H}_i}$ and $\Sigma_{n,{\cal H}_i}$  into $J_n$ in \eqref{g_J} we approximate 
$J_n\big(f(y_n|H_1),f(y_n|H_0)\big)$ 
as the following
\vspace{-1mm}
%
\begin{equation}\label{sim_j}
    J_n\big(f(y_n|H_1),f(y_n|H_0)\big) \!= \!\frac{\sigma_{w_n}^2\! + \!A_n g_n P_n}{\sigma_{w_n}^2 \!+ \!B_n g_n P_n}\!+\!\frac{\sigma_{w_n}^2 \! + \! C_n g_n P_n}{\sigma_{w_n}^2 \! + \! D_n g_n P_n}
\end{equation}
 where $A_n \!=\!P_{f_n}(1\!-\!P_{d_n}) + P_{d_n}(P_{d_n}\!-\!P_{f_n})$ and
\begin{align*} 
 C_n=&~P_{d_n}(1-P_{f_n}) - P_{f_n}(P_{d_n}-P_{f_n}),\nonumber \\
 B_n =& ~P_{d_n}(1-P_{d_n}), ~~D_n = P_{f_n}(1-P_{f_n}).
 \end{align*}
%
 \noindent Note that $J_n$ depends on the channel gain $g_n$ and power $P_n$.
 %
\section{Formalizing and Solving Optimal Transmit Power Adaptation  Problem} \label{XYZ}
\begin{figure}[!t]
	\centering
	\includegraphics[scale=.45]{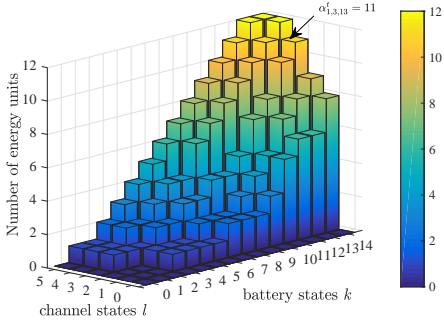}
	\caption{This numerical example shows how many energy units sensor $1$ should spend for transmitting its local decision $u_1=1$, given the knowledge of its channel and battery states. At slot $t$, if $g_1 \in {\cal I}_{1,3}$ and  the battery state is $b_1^t=13$ then $\alpha_{1,3,13}=11$. }
	\label{example}
	\vspace{-8mm}
\end{figure}
Recall $\alpha_{n,l,k}^t$ is the consumed units of energy for sensor $n$. Hence $p^t_{n,l,k} \! = \! \alpha^t_{n,l,k}e_u/T_s$ is the corresponding transmit power. Our main contribution is to design the {\it optimal transmit power map} with the points $p_{n,l,k}$ for $n = 1,\dots, N$, $l = 1,\dots, L$ and  $k = 1,\dots, K$. The optimal map can be found offline, via solving the following constrained optimization problem, and the map is shared with all sensors for distributed implementation. Sensor $n$ at slot $t$, given the knowledge of its quantized channel gain $l$ and its battery state $k$, decides which transmit power it should choose to transmit $u_n=1$.  Fig. \ref{example} shows an example of the optimal transmit power map for sensor $1$ when $L=5$ and $K=14$. Given our system model, we wish to maximize $J$-divergence subject to three constraints: the causality constraint, the battery outage constraint, and the total average power constraint. We characterize these three constraints in the following.
 First, the causality constraint \cite{Ozel} restrains the transmit power, such that the consumed energy for data transmission is less than the available energy in the battery, i.e. ${T_s} p_{n,l,k}\leq {k e_u}$. 
%
%
Second, the battery outage constraint prevents the sensor battery to be completely discharged. We express the battery outage constraint as
 \begin{align}\label{Const2}
\Pr\big (b_n^{t+1}> \eta b_n^t \; |\; b_n^t, l \big )\geq\zeta_n
\end{align}
 \noindent where $0\leq\eta \leq 1$. The battery outage constraint in \eqref{Const2} means that, with a probability of larger than $\zeta_n$, the energy units at slot $t+1$ will be larger than $\eta$-percent of the available energy units at slot $t$, given the channel state $l$.  By approximating $E_n \approx \beta_n e_u$, the constraint in \eqref{Const2} can be written as 
\begin{align}\label{sim_ii}
\Pr(b_n^{t+1}> \eta b_n^t \; | \; b_n^t, l) = &\Pi_0 F_{E_n}\big{(} k e_u (\eta \!-\!1)\big{)}  \nonumber\\
      + &\Pi_1 F_{E_n}\big{(}k e_u(\eta\!-\!1)+\alpha_{n,l,k}^t\big{)} 
\end{align}
 From \eqref{Const2} and \eqref{sim_ii} we find below
\begin{equation}
     \Pi_1F_{E_n} \Big{(}k e_u(\eta-1)+\alpha^t_{n,l,k}\Big{)}\leq 1-\zeta_n.
\end{equation}
Using the battery outage constraint we can find an upper bound on $p_{n,l,k}$, i.e., $p_{n,l,k}\leq \Phi_{n,k}$.
Using the CDF $F_{E_n}(\cdot)$, we find $\Phi_{n,k}$ as
%
\begin{align}
\Phi_{n,k} = -\frac{\gamma_e}{T_s e_u}\ln \left (\frac{ \Pi_1-1+\zeta_n}{ \Pi_1 }\right)-\frac{k  }{T_s}(\eta-1).
\end{align}
The total average power constraint for a given ${\cal P}_{tot}$ requires 
\begin{equation}
\sum_{n=1}^N\mathbb{E} [ P_n] \leq {\cal P}_{tot}
\end{equation}
 where $\mathbb{E} [ P_n] \! = \! \sum^L_{l=0}\sum^K_{k=0} p_{n,l,k} \pi_{n,l} \psi_{n,k}$. Considering the above three constraints, we have
 \vspace{-1mm}
 \begin{align}\label{max_j}
&\max_{\forall p_{n,l,k}} ~~\sum^N_{n=1}\sum^L_{l=0}\sum^K_{k=0} J_n(\mu_{n,l},p_{n,l,k})\pi_{n,l}\psi_{n,k}
    \nonumber\\
     &{\hbox{s.t. }}~(i)~~~p_{n,l,k}\leq \frac{k e_u}{T_s}\nonumber\\ &\qquad (ii)~~p_{n,l,k}\leq \Phi_{n,k} \nonumber\\ 
     &\qquad (iii)~\sum^N_{n=1}\sum^L_{l=0}\sum^K_{k=0}p_{n,l,k}\pi_{n,l}\psi_{n,k} \leq {\cal P}_{tot}
 \end{align}
where $J_n(\mu_{n,l},p_{n,l,k})$ in the cost function is the $J_n$ expression in \eqref{sim_j}, when $g_n$ and $P_n$, respectively, are replaced with $\mu_{n,l}$ (the corresponding quantized channel gain) and $p_{n,l,k}$ (the transmit power corresponding to channel state $l$ and battery state $k$). The optimization variables are $N\times L\times K$ points of the power map consisting of points $p_{n,l,k}$. Our system model can be viewed as an extension of the model in \cite{vin} in two aspects. We note that the system model in \cite{vin} does not include energy harvesting. Moreover, the developed  $J$-divergence optimal power allocation is based on perfect channel state information at the sensors. 
 The convexity of $J$-divergence function with respect to transmit powers is studied in \cite{vin} in terms of the local detection performance indices $0<P_{f_n}, \; P_{d_n}<1$. We state  the result in the following theorem.\vspace{-.1mm}
\par \textbf{Theorem 1 \cite{vin} :} The $J$-divergence optimization problem is convex when $(P_{d_n},P_{f_n})\in \mathcal{S}$ where the set $\mathcal{S}$ is
\begin{align*} 
\mathcal{S}= \Big \{ (P_{d_n},P_{f_n})~ \Big | ~\frac {3}{4} -\frac {1}{2}P_{f_n}\!- \!\frac {1}{4}\sqrt {1+12P_{f_n}-12P_{f_n}^{2}} \notag \\
\qquad \qquad \leq P_{d_n} \leq \frac {3}{4} -\frac {1}{2}P_{f_n}\! + \! \frac {1}{4}\sqrt {1+12P_{f_n}-12P_{f_n}^{2}}, \notag \\
\qquad \qquad \qquad \qquad \qquad  \qquad \qquad ~~~ 0 < P_{f_n} < P_{d_n} <1 \Big \}. 
\end{align*}
We assume that $P_{d_n}$ and $P_{f_n}$ in our system are such that $(P_{d_n},P_{f_n}) \in \mathcal{S}$ and hence the problem in (\ref{max_j}) is convex with respect to our optimization variables. 
%
Given the probabilities of the battery states $\psi_{n,k}$, we derive the solution of the problem in \eqref{max_j}, using the technique of Lagrange multipliers. The associated Lagrangian is
\begin{align*}
{\cal L}
\!=\!- \!\sum_{n=1}^N \! \sum^L_{l=0} \!\sum^K_{k=0}& \!\left[\frac{\sigma_{w_n}^2\!+\!A_n \mu_{n,l} p_{n,l,k}}{\sigma_{w_n}^2\!+\!B_n~\mu_{n,l}p_{n,l,k}}\!+\!\frac{\sigma_{w_n}^2\!+\! C_n \mu_{n,l} p_{n,l,k}}{\sigma_{w_n}^2\!+\!D_n\mu_{n,l}p_{n,l,k}}\right]
\nonumber\\ \times \pi_{n,l}\psi_{n,k} +&\lambda \left(\sum^N_{n=1}\sum^L_{l=0}\sum^K_{k=0}p_{n,l,k}\pi_{n,l}\psi_{n,k} - {\cal P}_{tot}\right).
 \end{align*}
 The Lagrangian multiplier $\lambda$ can be obtained using the ellipsoid method.
By setting $\partial {\cal L}/\partial p_{n,l,k} = 0$ we get
\begin{equation}\label{Lagran}
    -\frac {\left ({A_n - B_n }\right ) \sigma_{w_n} ^{2} \mu_{n,l}}{\left ({\sigma_{w_n} ^{2} \!+ \!B_n\mu_{n,l} p_{n,l,k} }\right )^{2}} \!-\! \frac {\left ({C_n - D_n }\right ) \sigma_{w_n}^{2} \mu_{n,l}}{\left ({\sigma_{w_n} ^{2} \!+\! D_n \mu_{n,l} p_{n,l,k} }\right )^{2}} \! + \!\lambda = 0.
\end{equation}
Suppose $p_{n,l,k}'$ is the solution to \eqref{Lagran} and $p_{n,l,k}^\ast$ denotes the solution to the problem in (\ref{max_j}). We have
\begin{equation} \label{solution}
p^{\ast }_{n,l,k} = \min \left \{{\frac{k e_u }{T_s} , \; \Phi_{n,k} , \; [p_{n,l,k}']^{+} }\right \},
\end{equation}
 and $\lambda$ satisfies the following 
 %
\begin{equation}
\lambda \left(\sum^N_{n=1}\sum^L_{l=0}\sum^K_{k=0}p^*_{n,l,k}\pi_{n,l}\psi_{n,k} - {\cal P}_{tot}\right)=0.
\end{equation}
%
 The corresponding optimal number of energy units for transmission is 
\begin{equation}\label{alpha_Opt}
\alpha^{\ast }_{n,l,k} = \lceil p^{\ast }_{n,l,k} T_s/e_u \rceil 
\end{equation}
\par
\begin{figure}[!t]
    \subfigure[ $\gamma_e = 0.5$]
    {        \includegraphics[scale=.2777]{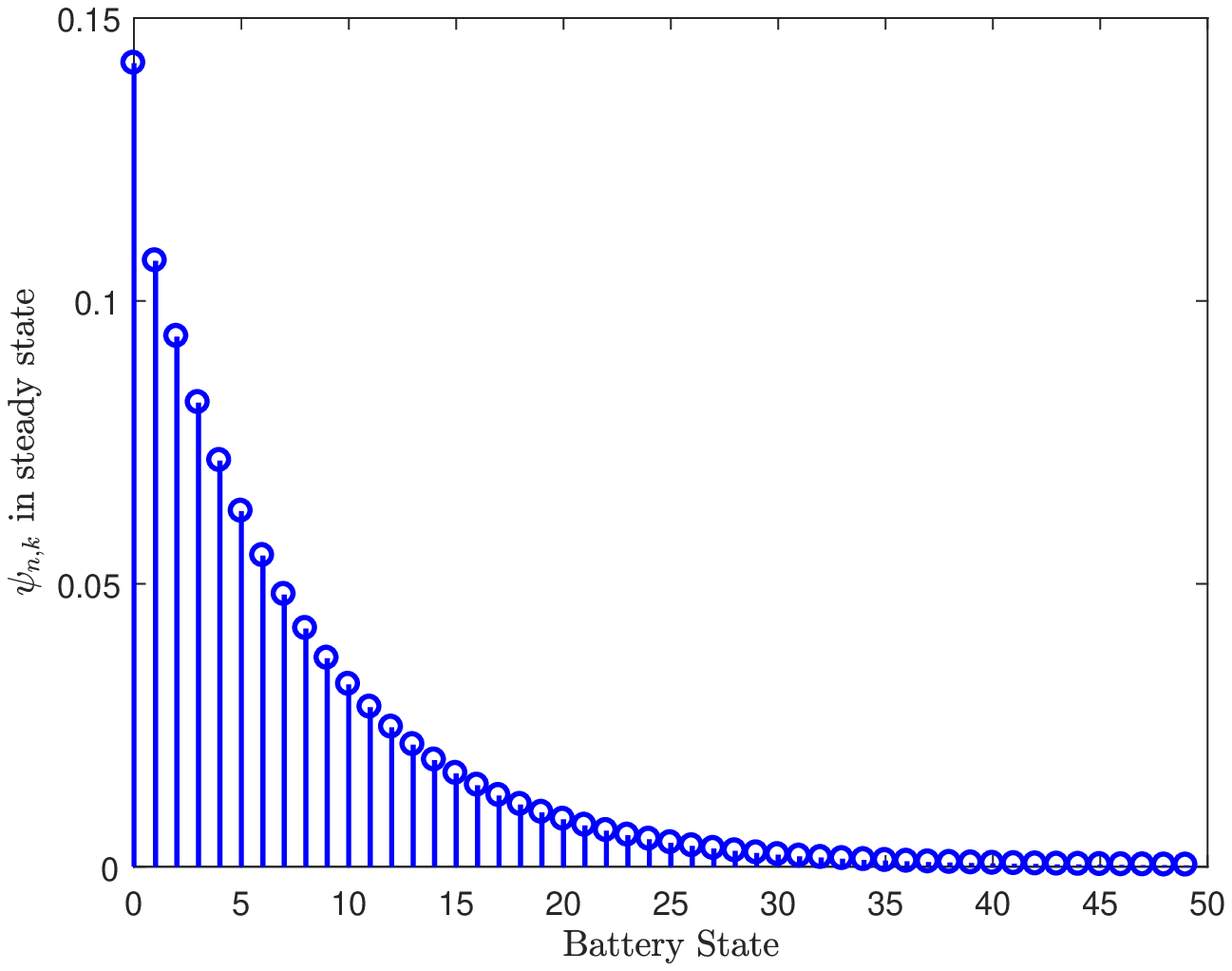}\label{fig2a}    }
    \subfigure[$\gamma_e = 1.5$]
    {        \includegraphics[scale=.2777]{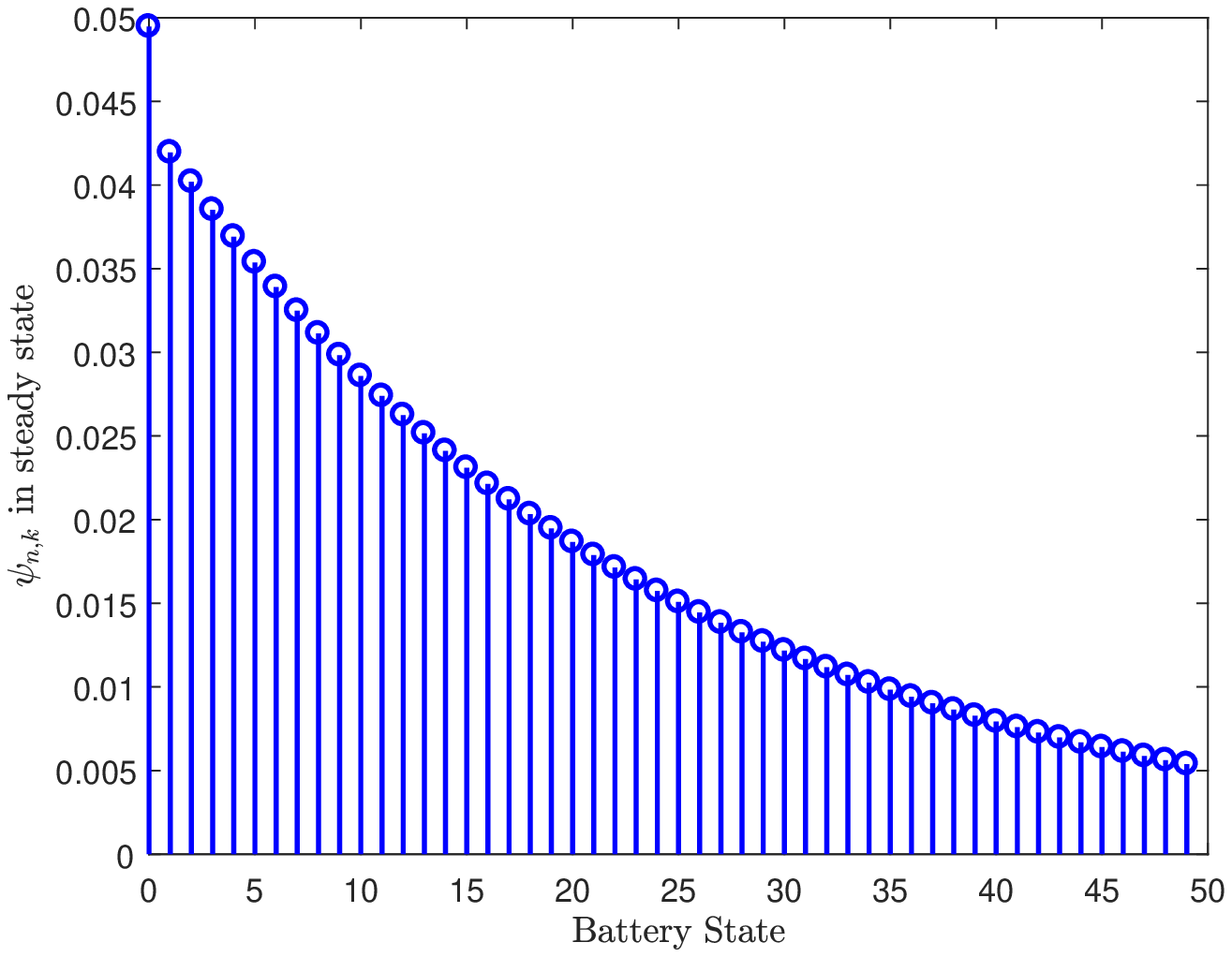}\label{fig2b}    }
    \caption{$\boldsymbol{\Psi}_n$ in the steady state for $K=50$.}
    \label{steadystate}
    \vspace{-2mm}
\end{figure}

 We take the following iterative approach to find the probabilities of the battery states. We start from an initial step (slot $0$) where the battery is fully charged, i.e. $\boldsymbol{\Psi} ^{(0)}_n =(\psi_{n,0}^0,\psi_{n,1}^0,\dots,\psi_{n,k}^0) = (0,0,\dots,1)$. We obtain $\alpha^{(0)}_{n,l,k}$ using 
Algorithm \ref{FingPower}. Then, we calculate $\boldsymbol{\Psi} ^{(1)}_n, \forall n$ using \eqref{prob_battery}. We iteratively find $\boldsymbol{\Psi} ^{(t)}_n$ and $\alpha^{(t)}_{n,l,k}$ until the convergence is reached, i.e., when the following criteria is met 
%
\begin{equation*}
\underset{k}{\max} \;|\psi_{n,k}^t -\psi_{n,k}^{t+1}|<\epsilon_2, ~~~\forall n.
\end{equation*}
%
The pseudo-code for calculating $\boldsymbol{\Psi}_n$ in the steady state is given in Algorithm. \ref{Alg1}. Fig.~\ref{steadystate} depicts $\boldsymbol{\Psi}_n$ in the steady state for different energy arrival rates $\gamma_e = 0.5, 1.5$. As $\gamma_e$ increases, the amount of harvested energy increases and thus the probability of the battery being discharged decreases. The performance of the proposed algorithm, will be illustrated and discussed next. \vspace{-.5mm}
%
 \begin{algorithm}[!b]
 \caption{ Finding $\alpha_{n,l,k}^{*(t)}$ given $\boldsymbol{\Psi}^{(t)}_n$}
 \label{FingPower}
 1: Input: $\boldsymbol{\Psi}^{(t)}_n$ \\
 2: Choose $\lambda^{(0)}$ and $t_0$, specify $\epsilon_1>0$, and set $i=0$.\\
 3: Calculate $p_{n,l,k}^{\prime(0)}$ by solving \eqref{Lagran}. \\
 4: Calculate optimal $p_{n,l,k}^{*(0)}$  using \eqref{solution}. \\
 \While {$\lambda^{(i)}\big( \sum_n\mathbb{E}[P_n^{*(i)}]-{\cal P}_{tot} \big)> \epsilon_1$,
 }
 {1:  $\lambda^{(i+1)}=[\lambda^{(i)}+ t_0 \big (\sum_n \mathbb{E} [P_n^{* (i)} ] -{\cal P}_ {tot}\big)]^+$.\\
 2: Calculate optimal $p_{n,l,k}^{\prime(i+1)}$ by solving \eqref{Lagran}. \\
 3:  Calculate optimal $p_{n,l,k}^{*(i+1)}$ using \eqref{solution}. \\
 4: Set $i = i + 1$.
 }
 5: Calculate optimal $\alpha_{n,l,k}^{*(t)}$ using \eqref{alpha_Opt}.
 \end{algorithm}
 \begin{algorithm}[!b]
 \caption{Finding $\boldsymbol{ \Psi}_n$ in the steady state}
 \label{Alg1}
 1: Specify $\epsilon_2>0$, set $\boldsymbol{\Psi} ^{(0)}_n= (0, 0,\dots,1)$ and $q = 0$ \\
 2: Calculate optimal $\alpha_{n,l,k}^{*(0)}$ using Algorithm.\ref{FingPower}. \\
 3: Update $\boldsymbol{ \Psi}^{( 1)}_n$ by using \eqref{prob_battery}. \\
 \While {$\underset{k}{\max} \;|\psi_{n,k}^q -\psi_{n,k} ^{q+1}| > \epsilon_2, \forall n$,
 }
 {1: Calculate optimal $\alpha_{n,l,k}^{*(q+1)}$ using Algorithm.\ref{FingPower}.\\
 2: Update $\boldsymbol{ \Psi}^{( q+1)}_n$  using \eqref{prob_battery}.\\
 3: Set $q = q + 1$.
 }
 \end{algorithm}

 \section{simulation results}\label{sim}

 In this section, we provide the numerical results to illustrate our proposed power adaptation scheme. Our simulation parameters are  $N\!=\!2$, $\gamma_{g_n}=[1.1, 1.2]$, $K=100$, $\sigma_{w_n}^2=[1, 1.5]$, $\mu = [0,  0.1,  0.3,  0.6,  1.2,  \infty]$ for all sensors, $P_{f_n} = [0.2, 0.1]$ and $P_{d_n} = [0.9, 0.75]$. Note that sensors are heterogeneous, in the sense that their statistical information parameters are different. We set $e_u = 0.1$ Joules, $T_s = 0.1$ s, $\zeta_n = 0.9$ and $\eta = 0.2$ for all sensors. The FC uses a Neyman-Pearson detector based on the likelihood ratio of the received signal $\boldsymbol{y}$. Let $P_D^{FC}$ and $P_F^{FC}$, respectively, represent the detection and false alarm probabilities of NP detector at the FC. The threshold of the NP detector is determined by the target $P_F^{FC}= 0.1$. Assuming sensors use the optimal power map from section \ref{XYZ} to adapt their transmit powers when sending their local decisions $u_n=1$, we find $P_{D}^{FC}$.
\begin{figure}[!t]
	\centering
	\includegraphics[scale=.45]{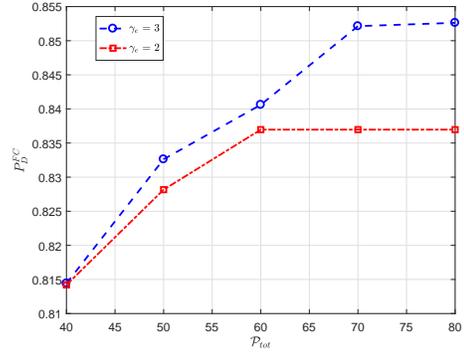}
	\caption{$P_{D}^{FC}$ vs. $P_{tot},~ K=100, P_F^{FC}=0.1$}
	\label{PD1}
	\vspace{-3mm}
\end{figure}
%
\begin{figure}[!t]
	\centering
	\includegraphics[scale=.45]{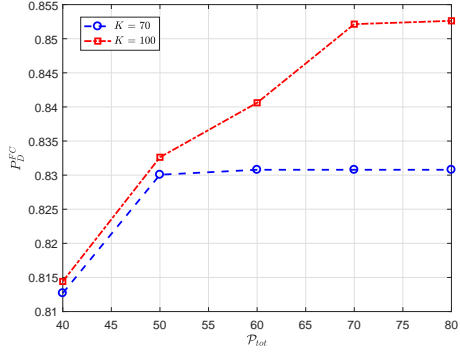}
	\caption{$P_{D}^{FC}$ vs. $P_{tot},~\gamma_e=3, P_F^{FC}=0.1$}
	\label{PD2}
	\vspace{-3mm}
\end{figure}
%
\par
\begin{figure}[!t]
	\centering
	\includegraphics[scale=.45]{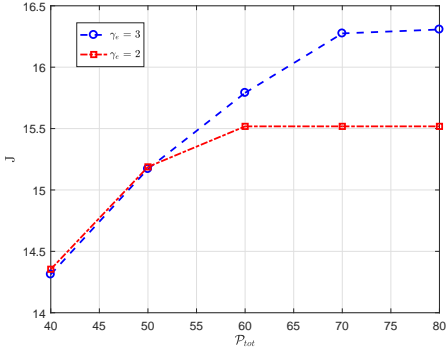}
	\caption{$J$ vs. $P_{tot},~ K=100$}
	\label{J1}
	\vspace{-3mm}
\end{figure}

\begin{figure}[!t]
	\centering
	\includegraphics[scale=.45]{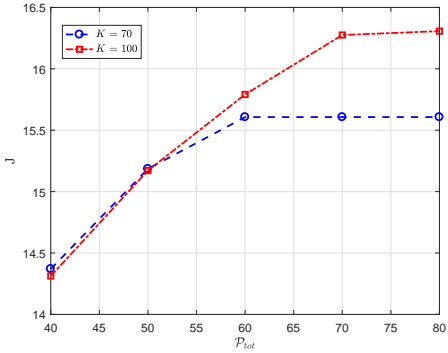}
	\caption{$J$ vs. $P_{tot},~\gamma_e = 3$}
	\label{J2}
	\vspace{-4mm}
\end{figure}

 Fig.~\ref{PD1} and Fig.~\ref{PD2} show $P_{D}^{FC}$ versus ${\cal P}_{tot}$. Note that $P_{D}^{FC}$ increases as ${\cal P}_{tot}$ increases, however, it remains almost the same after ${\cal P}_{tot}$ reaches and exceeds a certain value. This is due to the fact that (depending the battery capacity $K$ and the energy arrival rate $\gamma_e$), the total power constraint in (\ref{max_j}) becomes and remains inactive when ${\cal P}_{tot}$ reaches and exceeds this certain value. In this case, the optimal $\lambda$ becomes zero and the sensors' transmit power $P_n$'s do not change. Fig.~\ref{PD1} shows $P_{D}^{FC}$ versus ${\cal P}_{tot}$ for $\gamma_e = 2, 3$. As $\gamma_e$ increases, the saturation of $P_{D}^{FC}$ occurs at a larger value of ${\cal P}_{tot}$.
 Fig.~\ref{PD2} shows $P_{D}^{FC}$ versus ${\cal P}_{tot}$ for $K = 70, 100$. As $K$ increases, the saturation of $P_{D}^{FC}$ happens at a larger value of ${\cal P}_{tot}$. Fig.~\ref{J1} and Fig.~\ref{J2} show $J$-divergence versus ${\cal P}_{tot}$, confirming that it has the same trend as $P_{D}^{FC}$ versus ${\cal P}_{tot}$.
   %

%

\section{Conclusions and Future Research}\label{conclu}

In summary, we studied optimal transmit power adaptation for binary distributed detection problem in a WSN with heterogeneous energy harvesting sensors. Aiming at maximizing the approximate $J$-divergence of the received signal densities under two hypotheses at the FC (subject to certain constraints), we provided the optimal power map, which would become available at the sensors. The optimal power map depends on the energy arrival rate, the battery capacity, and the battery states probabilities at the steady state. Using the optimal power map, each sensor chooses its transmit power, based on its battery state and its channel state. Through simulations, we investigated the performance of our proposed power adaptation scheme for different system parameters. 
For future research, we expand our system model and in particular the wireless communication channel model, and consider a finite-state Markovian fading channel model. We will study the performance of our proposed power adaptation scheme and compare it with other power allocation methods in the literature for energy harvesting systems, including dynamic programming techniques.
\section*{Acknowledgment}
This research is supported by NSF under grants CIF-1341966 and CIF-1319770.
\bibliographystyle{IEEEtran}
\bibliography{RefEH}
\end{document}